\documentclass[letter, 10pt, conference]{ieeeconf}  

\usepackage{graphics} 
\usepackage{graphicx}\graphicspath{ {./figures/} }
\usepackage{amsmath} 
\usepackage{bm}
\usepackage{multicol}
\usepackage{booktabs}
\usepackage[linesnumbered,ruled]{algorithm2e}
\usepackage[font={small}]{caption}
\usepackage{subcaption}
\usepackage{mathtools}
\usepackage{wrapfig}
\DeclareMathOperator*{\argmin}{arg\,min}
\DeclareMathOperator*{\argmax}{arg\,max}
\usepackage{url}
\usepackage[para]{footmisc}

\title{\LARGE \bf
A behavior driven approach for sampling rare event situations for autonomous vehicles.
}

\author{Atrisha Sarkar and
        Krzysztof Czarnecki\\
University of Waterloo.\\
Canada\\
Email: atrisha.sarkar@uwaterloo.ca, kczarnec@gsd.uwaterloo.ca
}

\begin{document}

\maketitle

\begin{abstract}

Performance evaluation of urban autonomous vehicles requires a realistic model of the behavior of other road users in the environment. Learning such models from data involves collecting naturalistic data of real-world human behavior. In many cases, acquisition of this data can be prohibitively expensive or intrusive. Additionally, the available data often contain only typical behaviors and exclude behaviors that are classified as rare events. To evaluate the performance of AV in such situations, we develop a model of traffic behavior based on the theory of bounded rationality. Based on the experiments performed on a large naturalistic driving data, we show that the developed model can be applied to estimate probability of rare events, as well as to generate new traffic situations.
\end{abstract}

\section{INTRODUCTION}

With autonomous vehicles (AV) poised to change the transportation landscape, the ability of AVs to handle a wide range of human traffic behaviors safely and reliably is of paramount importance. In order to guarantee that, it is inadequate to rely on field tests alone as a primary method of AV evaluation, since the number of kilometers that needs to be driven for any statistical safety guarantee is prohibitively high \cite{kalra2016driving}. Thus, there is an increasing role of simulation in all major components of an autonomous driving system (ADS), including perception, planning, testing, and verification. Although it is possible to significantly speed up the verification process in simulation, it is also necessary for simulation environments to be realistic. For the behavior planner (which is the component in ADS responsible for tactical and high-level decision making), this means that simulation environments should be able to model behavior of other traffic users in a way that is reflective of the real-world behavior. Popular approaches design this behavior in several ways: \emph{expert-driven}, where designers program the motion and behavior of the users \cite{yuan2015agent}, \emph{data-driven} where a model of behavior is learnt from observations and naturalistic driving datasets \cite{chen2010calibration}, or a \emph{hybrid} model that uses a combination of both \cite{zeng2014application}\cite{sarkar2017trajectory}. Although it is possible to design models that learn from real-world data, a major challenge in any approach is the generation of unusual or atypical behavior that is not readily observed in the data, such as crashes or near-miss scenarios. \\

In dynamical systems, rare event (RE) sampling provides a mathematical framework to analyse events of very low probability \cite{rubino2009rare}. RE sampling techniques can be used for both estimating the probability of occurrence of rare events, as well as generating the conditions that lead to rare events. In recent years, RE sampling based techniques have been used for simulation based verification and testing of a wide range of motion and behavior planners. O'Kelly et al. use RE sampling for testing of planners that work in end-to-end manner based on deep learning \cite{o2018scalable}, whereas, other approaches apply similar techniques to evaluate performance in specific traffic situations, such as lane changes and cut-ins \cite{zhao2017accelerated}. Most approaches that use rare event sampling for AV evaluation, uses cross-entropy based importance sampling, which is an adaptive sampling technique to search for a sampling distribution that maximizes odds of leading to crashes and near-miss scenarios.\\
\\
A part of the uncertainty in traffic environments arises from the inherent stochastic behavior of road users, as reflected in different driving styles of human drivers. This is in contrast to the design of motion and behavior planners of an ADS, which optimize a set of defined objectives, such as, progress, safety, observance of traffic rules, etc. One way to bridge this disconnect is by applying the theory of \emph{bounded rationality}. Originally developed in the field of behavioral economics to study decision making of rational agents, the theory of bounded rationality is a standard technique in artificial intelligence that provides a general framework to model sub-optimality of human decision making behavior \cite{simon1972theories}. In this work, we apply bounded rationality to model stochastic traffic behavior, and develop a novel rare event sampling and optimization mechanism that provides greater interpretability to the generation of rare event situations than the standard cross-entropy approach. We show that categorizing different driving behaviors, and optimizing for an appropriate driving policy can act as an effective technique for rare event estimation. We compare our approach to cross-entropy based importance sampling, which is a standard technique for rare event estimation. We show that our method has lower variance, thus provides a better estimate of rare event probability, as well as provides 33$\%$ speed-up over cross entropy sampling and a speed-up to the order of $10^4$ compared to crude Monte Carlo sampling. Further, we fit the behavior model to a naturalistic driving dataset, and evaluate its use for the generation of new situations. Our evaluation is based on vehicle cut-in events from the University of Michigan SPMD naturalistic driving dataset, which contains several hours of real world driving.\\

\section{BACKGROUND}

\subsection{Rare event sampling}
\label{sec:re_sampling}

Rare event sampling provides a framework to study events of very low probability in dynamical systems, with the primary goal of estimating their probability. In a dynamical system, the generalized system dynamics can be expressed as
\begin{equation*}
X_{t+1} = \Phi(X_{t},G_{t})
\end{equation*}

where $X_{t} \in \mathbf{R}^{n},G_{t} \in \mathbf{R}^{m}$ are the system state and input at time $t$, and $\Phi$ is the system dynamics. A rare event $\epsilon$ is defined using a scalar performance function $\eta$, and the rare event is the occurrence of the condition where the function is greater or equal to a specified threshold, $\eta(X_{t}) \geq b$. For convenience, rare events can also be defined over the input space, i.e, the set of inputs that leads the system trajectory to a critical state as defined by $\eta(\cdot)$ and $b$. Thus, the probability of rare events can be expressed by the following integral

\begin{equation} 
p_{\epsilon} = \int_{\epsilon} I_{\epsilon}(g)p(g)dg
\label{eqn:intgl_pe}
\end{equation}

where $I_{\epsilon}(g)$ is the indicator function, such that, $I_{\epsilon}(g)=1$ if $g \in \epsilon$, and $I_{\epsilon}(g)=0$ otherwise. In crude Monte Carlo methods, the above integral can be approximated by generating several independent and identically distributed samples\footnote{we use the convention where small $g$ is the random variable and capital $G$ is a sample.} of the system inputs $G^{(0)},G^{(1)},..,G^{(N)}$ drawn from the distribution $p(g)$ as

\begin{equation}
    p_{\epsilon} \approx p_{\epsilon}^{MCS} = \frac{1}{N} \sum_{i=1}^{N}{I_{\epsilon}(G^{(i)})} 
\label{eqn:int_aprox}
\end{equation}
By the Central Limit theorem, as $N \rightarrow \infty$, $p_{\epsilon}^{MCS}$ is distributed asymptotically as a Gaussian distribution with mean $p_{\epsilon}$ and variance $\sigma^{2}=$   $\frac{p_{\epsilon}(1-p_{\epsilon})}{N}$. Since the above is only an estimate of the true probability, it is important to calculate the margin of error based on the number of samples. The relation between the sample size and relative margin of error ($re$) of estimation for a confidence interval of 0.95 is given by the relation\footnote{Derivation in \cite{rubino2009rare} page 4.} $N > \frac{1.96}{re^{2}p_{\epsilon}}$. Thus, even for a relative error of .01 and a high rare event probability of .001, we need $10^7$ samples.\\
Importance sampling is a variance reduction technique that helps in improving the accuracy of the estimate $p_{\epsilon}$ with fewer samples. A major disadvantage of crude Monte Carlo sampling is the variance of the samples in \eqref{eqn:int_aprox}, where for rare events, most $I_{\epsilon}(G^{(i)})$s are 0 and only very few are 1. To reduce this variance, instead of drawing samples from $p(g)$, importance sampling uses a proposal distribution $q(g)$, where samples drawn from $q(g)$ have a higher probability of leading to rare event situations. Using $q(g)$, the integral in \eqref{eqn:intgl_pe} can be written as
\begin{equation*} 
p_{\epsilon} = \int_{\epsilon} \frac{I_{\epsilon}(g)p(g)}{q(g)}q(g)dg
\end{equation*}
which is the expectation $\mathbf{E}_{q}[\frac{I_{\epsilon}p}{q}]$. Following a similar approach to crude Monte Carlo, the above integral can be approximated as
\begin{equation*} 
p_{\epsilon} \approx p_{\epsilon}^{IS} =  \frac{1}{N} \sum_{i=1}^{N}{I_{\epsilon}(G^{(i)})w(G^{(i)})} 
\end{equation*}
where $w(G^{(i)})=\frac{p(G^{(i)})}{q(^{(i)})}$ is the \emph{importance} weight of the sample $G^{(i)}$ or the \emph{likelihood ratio}. The above estimator is an unbiased estimator with $\mu = p_{\epsilon}$ and $\sigma^{2} = \frac{1}{N}(\mathbf{E}_{q}[\frac{I_{\epsilon}p^{2}}{q^{2}}] - p_{\epsilon}^{2})$ 
 under the condition that when $I(g)p(g)>0$, $q(g)>0$ holds true. Thus, for every sample drawn from $p(\cdot)$ that leads to a rare event, the same sample should also lead to a rare event if they were drawn from $q(\cdot)$. The complete algorithm to estimate $p_{\epsilon}$ based on IS technique is shown in Algorithm \ref{algo:is}.

The most optimal choice for the proposal distribution $q(\cdot)$ is the original distribution conditioned on the rare event $\epsilon$
\begin{equation}
q^{*}(\cdot) = p(g|\epsilon) = \frac{I_{\epsilon}(g)p(g)}{p_{\epsilon}}
\label{eqn:opt_q}
\end{equation}
We can see that with this choice the variance reduces to zero. Additionally, since the distribution is conditioned on $\epsilon$, any sample drawn from this distribution would have $I_{\epsilon}$ value of 1 and weight $w = p_{\epsilon}$. Thus, just a single sample can estimate the exact value of $p_{\epsilon}$.  In most cases, it is practically impossible to find the exact distribution $p(g|\epsilon)$ since it requires knowing $p_{\epsilon}$. However, this optimal distribution gives an indication that a distribution close to $p(g|\epsilon)$ is a good proposal distribution.\\
One approach to generate the proposal distribution is based on cross-entropy (CE) method. If $q$ is chosen from a family of distributions $\psi(\cdot,\theta)$, then the distance between the distribution $\psi(\cdot,\theta)$ and $p(g|\epsilon)$, as measured by Kullbeck-Leibler divergence, gives an estimate of goodness of the proposal distribution. Thus, the optimal distribution can be found by solving the following optimization problem:
\begin{equation}
    \theta^{*} = \argmin_{\theta} D_{KL}(\psi(\cdot,\theta),p(g|\epsilon))
    \label{eqn:ce_min}
\end{equation}
where $D_{KL}$ is the KL divergence between the two distributions, and $\theta^*$ is the parameter of the optimal distribution. The CE approach gives a fast iterative scheme to find the solution to the above minimization problem. 
The optimization procedure can be further simplified by choosing $\psi(\cdot,\theta)$ as the exponentially change of measure (ECM) function of the original distribution $p(g)$. Most approaches for rare event sampling in AV literature approximate $p(g)$ with a heavy-tailed distribution from the exponential family.
This enables CE optimization to find a closed form solution to \eqref{eqn:ce_min}, thus making the process much faster.\\
Although CE approach provides a point estimate of the probability of rare events, it provides little insight into the type of behaviors that lead to the rare event situations. This becomes especially relevant for evaluation of AVs, where it is important to know what conditions and behavior cause rare event situations.  

\begin{algorithm}
\SetAlgoLined
\KwResult{IS estimate $p_{\epsilon} = \frac{1}{N}\cdot\sum\limits_{i=1}^{N} w_{i}$}
 \ForEach{$i \in [0,..,N]$}{%
      Sample an initial condition from the proposal distribution $g_{i} \sim q(g)$ \;
      Compute system trajectory based on the system dynamics : $\mathbf{X_{i}}={X_{t=0},X_{t=1},...,X_{t=T}}$\;
      \If{$\max\limits_{0 \leq i \leq T}$ g($\mathbf{X_{i}}) \geq b$}{%
        \textup{Calculate the importance weight of the $i^{th}$ particle} $w_{i}=\frac{p(g_{i})}{q(g_{i})}$ \;
        }
      }
 \label{algo:is}
 \caption{Importance sampling algorithm for estimating probability of rare events.}
\end{algorithm}
\subsection{Bounded rationality}
\label{sec:br}

Utility-based models are a widely used framework for motion planning in AVs. These models generate a driving policy by optimizing a set of \emph{utility functions} that typically include objectives such as safety, progress towards a goal destination, abiding by regulatory traffic rules, etc. Although a pure utility maximization principle can generate a driving policy for the subject vehicle, the approach has limited ability to capture the variation in behavior of other traffic users \cite{schildbach2015scenario}, \cite{wang2018reinforcement}. This shortcoming is mainly due to the fact that natural human behavior is often \emph{sub-optimal}, as they have limited ability to calculate the most optimal decision with respect to a set of utility objectives. The theory of \emph{bounded rationality} is a standard technique in artificial intelligence that provides a formal framework to model this sub-optimality of human behavior \cite{wright2010beyond}. One way to model bounded rationality is through the \emph{quantal response function}, which in its basic form gives the probability $P(s,A_i|\lambda)$ of a taking a discrete action $A_i \in \mathbf{A}$ in environment state $s \in S$ based on a utility $u: S \times A \rightarrow [-1,1]$ as
\begin{equation}
    P(s,A_i|\lambda) = \frac{\exp[\lambda \cdot u(s,A_i)]}{\sum\limits_{\forall A}\exp[\lambda \cdot u(s,A)]}
    \label{eqn:qr}
\end{equation}
$\lambda$ is the \emph{rationality parameter}, that controls the probability of an action $A_i$ based on its utility $u(s,A_i)$. When $\lambda \rightarrow \infty$, the policy converges to a pure utility maximization, i.e., the agent always takes the optimal action. Whereas, $\lambda \rightarrow 0$ leads to a random policy. 
However, since most actions in the context of an ADS are continuous (such as target velocity, distance to other vehicles, etc.), first we extend the quantal response function of \eqref{eqn:qr} to continuous actions. Second, since utilities for behavior planning in AVs are multi-objective in nature, we need to model the variation of behavior with respect to individual utilities. Thus, we extend the rationality parameter to a vector of tuples $\Lambda= [(\lambda_{1},u_{1}),(\lambda_{2},u_{2})..,(\lambda_{k},u_{k})]$, where each $\lambda_i$ acts as the rationality parameter for the corresponding utility $u_i$. With the above two extensions, we model the probability of an action $a \in \mathbf{R}^D$ in state $s$ by the probability distribution:
\begin{equation*}
      p(s,a|(\lambda,u)) = 
  \begin{dcases}
    \frac{1}{2} & \lambda=0 \\
    \frac{\lambda \cdot \exp(\lambda \cdot (1+u(s,a)))}{\exp(\lambda \cdot 2)-1}  & \lambda \neq 0 \\
  \end{dcases}
\end{equation*}
Combining the individual utilities, the final stochastic driving policy $f: S\times A \rightarrow [0,1]$ is formulated by the following mixture distribution:
\begin{equation}
f(s,a|\Lambda) = \frac{1}{k}\sum\limits_{i=1}^{k} p(s,a|(\lambda_{i},u_{i})) 
\label{eqn:drive_pol_f}
\end{equation}

\begin{figure}[!htbp]
\centering
\includegraphics[width=0.35\textwidth]{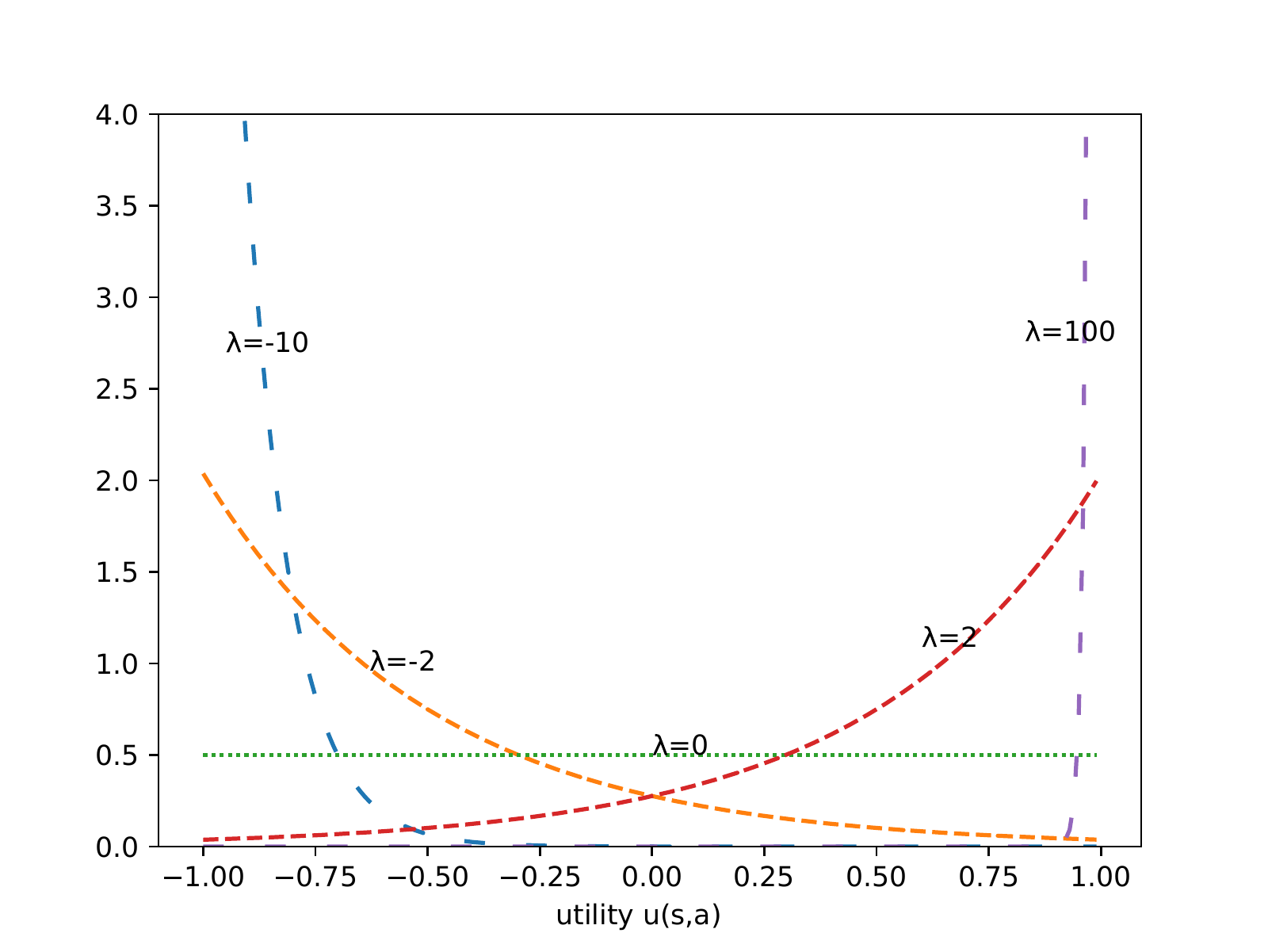}
\setlength{\belowcaptionskip}{-5pt}
\caption{Probability distribution of an action $a$ based on its utility $u(s,a)$. The plot shows the effect of the rationality parameter $\lambda$ on the probability (higher $\lambda$ leads to actions favoring higher utility).}
\label{fig:l_var}
\end{figure}

Figure \ref{fig:l_var} shows the relation between the action probability $p(s,a)$ and the action utility $u(s,a)$ at various levels of $\lambda$. A high value of $\lambda =100$, skews the distribution $p(s,a)$ such that the policy prefers actions with maximum utility ($u(s,a)=1$) with close to probability 1, i.e., a pure utility maximization model. Whereas, lower values of $\lambda=2,-2,-10$ progressively lead to a more sub-optimal policy. $\lambda=0$ being the special case where actions are chosen based on a random policy. As we discuss in the next section, the flexibility of the bounded rationality model afforded by the rationality vector $\Lambda$ helps to model a wide range of naturalistic driving behavior.  

\section{RARE EVENT SAMPLING AND SITUATION GENERATION}

\subsection{Behavior categorization}
\label{sec:beh_cat}

\begin{figure}[!htbp]
\centering
\includegraphics[scale=1]{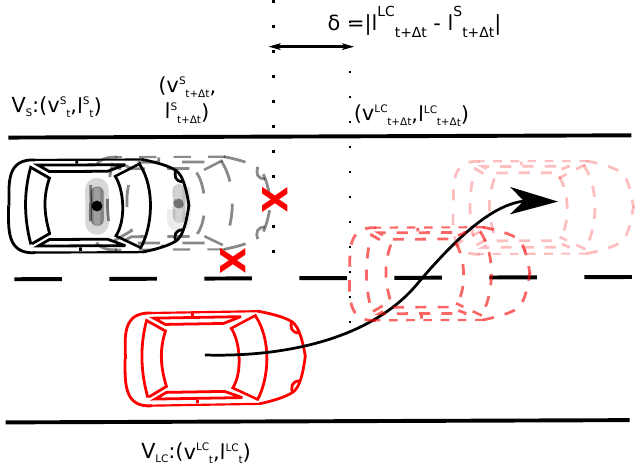}
\setlength{\belowcaptionskip}{-10pt}
\caption{Vehicle cut-in scenario on a two-lane road. $v^{LC,S}_{t}$ are the velocities of autonomous subject vehicle ($V_{S}$) and the target vehicle ($V_{LC}$) resp. at the start of the lane-change maneuver. $v^{\{LC,S\}}_{t+\Delta t}$ and $l^{\{LC,S\}}_{t+\Delta t}$ are the velocities and locations when the front wheel of the target crosses the lane boundary. $\delta$ is the distance gap.}
\label{fig:cut_in}
\end{figure}

In this section we use a typical vehicle cut-in scenario as a motivating example, and show how the bounded rationality model developed in the previous section can be used to categorize a range of driving behaviors. A vehicle cut-in scenario (Figure \ref{fig:cut_in}) involves a vehicle ($V_S$) maintaining its lane of travel, while another vehicle ($V_{LC}$) is executing a lane-change maneuver into $V_S$'s lane of travel. We consider the case where $V_{S}$ is driven in autonomous mode (subject vehicle), and $V_{LC}$ is driven by a human (target vehicle). There are two conflict points, as marked by a cross in the figure --- a side-to-side conflict that can result in sideways collision, and a sequential conflict that can result in rear-end collision.\\ 
\begin{table}[!htbp]
  \begin{center}
    \caption{Behavior categories based on the contraints on the rationality vector $\Lambda = [(\lambda_{\delta},.),(\lambda_{\tau},.),(\lambda_{p},.)]$}
    \label{tab:beh_cat}
    \begin{tabular}{p{1.4cm}lp{5cm}}
      \toprule 
      \textbf{Behavior id} & $[\lambda_{\delta},\lambda_{\tau},\lambda_{p}]$ & \textbf{Behavior description}\\
      \midrule 
      B1 & $[-,-,+]$ & cut-in with high speed at close distance with low ttc\\
      B2 & $[-,+,+]$ & high speed at close distance with high ttc\\
      B3 & $[+,+,-]$ & low speed at longer distance with high ttc\\
      B4 & $[+,-,-]$ & low speed at longer distance with low ttc\\
      B5 & $[-,-,-]$ & low speed at close distance with low ttc\\
      B6 & $[-,+,-]$ & low speed at close distance with high ttc\\
      B7 & $[+,+,+]$ & high speed at longer distance with high ttc\\
      B8 & $[+,-,+]$ & high speed at longer distance with low ttc\\
      \bottomrule 
    \end{tabular}
  \end{center}
\end{table}
As a part of the cut-in maneuver $V_{LC}$ needs to decide on its target velocity $v^{LC}_{t+\Delta t}$, where $t+\Delta t$ is the time step when the front wheel of $V_{LC}$ crosses the lane boundary of $V_S$'s travel lane. $V_{LC}$ also needs to decide the safety distance (distance gap) it has to keep from $V_S$, as measured by the difference between the vehicles' respective positions along the direction of travel $\delta = d(l^{LC}_{t+\Delta t},l^{S}_{t+\Delta t})$. Based on these choices, $V_{LC}$'s cut-in behavior can vary significantly. For example, $V_{LC}$ can cut-in close to $V_S$ with a low relative speed, or execute a high-speed maneuver maintaining a fair distance gap. Thus, the action space for $V_{LC}$ for the maneuver consists of the tuple $(v^{LC}_{t+\Delta t},\delta)$, and the state space consists of $v^{S}_t$. Once $V_{LC}$ initiates the cut-in maneuver, the subject vehicle $V_S$ needs to respond appropriately based on its behavioral decision logic,  which might include slowing down to maintain a safe distance gap or time-to-collision (ttc).\\
To model the behavior of the target vehicle $V_{LC}$ qualitatively, we use three utility functions; two based on safety ($u_\delta$,$u_\tau$), measured by the distance gap and time to collision (ttc), and one based on progress ($u_p$), measured by the velocity ($v^{LC}_{t+\Delta t}$). $u_{\{\delta,\tau\}}(x) = S(x-\{\delta^{*},\tau^{*}\}) + 0.5S(\{\delta^{*},\tau^{*}\}-x)$ and $u_p(v^{LC}_{t+\Delta t})=S(2v^{LC}_{t+\Delta t}-2v^{*}) - S(2v^{*}-2v^{LC}_{t+\Delta t})$, where $S$ is the standard logistic sigmoid function, $x$ is ttc or distance gap for the respective utilities, and $\delta^{*},\tau^{*},v^{*}$ are the parameter values that are based on safe driving best practices. These functions belong to the general class of exponential utility functions, which is a popular class of utility functions used in decision theory \cite{fishburn1970utility}. The choice is also based on insights from \cite{kageyama1992new}, which shows that a driver's perception of risk level and their response has an exponential relation to critical vehicle and environmental states, such as, getting close to an obstacle or curbside, and vehicle speed. Based on these utilities, we can construct the rationality vector $\Lambda =[(\lambda_{\delta},u_{\delta}),(\lambda_{\tau},u_{\tau}),(\lambda_{p},u_{p})]$. Following the equations in section \ref{sec:br}, every instance of the vector $\Lambda$ generates a stochastic driving policy $f(v_{S},(v^{LC}_{t+\Delta t},\delta))$, and $\lambda_{\delta},\lambda_{\tau},\lambda_{p}$ control the level adherence of the policy to each utility.\\
As shown in Table \ref{tab:beh_cat}, based on the level of adherence, $\lambda$'s in $\Lambda$ can be grouped together to form categories of driving behavior. For example, the constraint $\lambda_{\delta}<0,\lambda_{\tau}<0,\lambda_{p}>0$ leads to a driving policy that cares less about maintaining a safe distance-gap and time-to-collision, but more about making fast progress; shown in the table as the behavior category B1. Thus, we get eight behavior categories for a vehicle cut-in scenario, and even within a category there are a wide range of individual driving policies sharing the common behavior. We model the response of the subject vehicle $V_{S}$ based on Krauss car following model \cite{krauss1997metastable}, which is activated at time step $t+\Delta T$.
\subsection{Parameter optimization}

In this section, we develop an optimization scheme and show how the developed behavior model can be used for the purpose of rare event sampling. Revisiting \eqref{eqn:opt_q}, an optimal proposal distribution $q(\cdot)$ for importance sampling of rare events should be as close as possible to the distribution $p(g|\epsilon)$. In other words, the goal is to find a low variance estimator that has high probability in regions of the system input space $u$ that lead to rare events. One way to achieve that is by finding a driving policy that is more likely to lead to such events. To that end, we use the parameterized driving policy of \eqref{eqn:drive_pol_f} as the proposal distribution $q(\cdot)$. The system input space ($g$) is $\mathbf{R}_{>0}^2$ which consists of the velocity of the target vehicle and the distance gap. Thus, $q = f(g|\mathbf{\Lambda^{*}})$, where $\mathbf{\Lambda^{*}}$ is the solution to the following optimization problem

\begin{equation*}
 \mathbf{\Lambda^{*}} = \argmax_{\mathbf{\Lambda}} I_{\epsilon}(g)f(g|\mathbf{\Lambda})   
\end{equation*}

where $I_{\epsilon}(g)$ is the indicator function for the rare events. To solve the above optimization problem, we use a Simulated Annealing (SA) based heuristic that first finds the category of behavior (B1-8) that has a high $I_{\epsilon}(g)$ and then subsequently finds a value of $\mathbf{\Lambda}$ within that behavior category that maximizes the optimization objective.\\
Algorithm \ref{algo:br_opt} describes the optimization procedure. The two main structures in the algorithm, $[p_{max}^{B1},..,p_{max}^{B8}]$ and $[\Lambda_{max}^{B1},..,\Lambda_{max}^{B8}]$ maintain the maximum probability of rare event for each behavior category, and the corresponding $\Lambda$ of the driving policy that caused the rare events.\\
There are two loops in the procedure, the outer loop iterates over all behaviors to find a behavior with maximum rare event probability (line 5), and the inner loop iterates to find the $\Lambda$ that maximizes rare events within the behavior category line (10). Following the standard technique in Simulated Annealing, the acceptance of a better solution (line 8 and 14) is controlled by the temperature parameters ($T_{out},T_{inn}$), which are reduced by a constant factor in every iteration of the loop (line 18 and 21). The neighborhood generation of the outer loop (line 6) performs a weighted sampling of behavior ids based on the current $[p_{max}^{B1},..,p_{max}^{B8}]$ vector at each iteration.\\
For the inner loop, the \texttt{sample()} method generates a value of $\Lambda$ constrained by the behavior category based on a uniform distribution (line 11). \texttt{simulate\_scene()} is the main entry point to simulate a set of situations with different initial conditions. In our implementation, we use the SUMO open source simulator to simulate the cut-in scenarios \cite{krajzewicz2002sumo}. The method samples the initial state $v^{S}_{t}$ from the distribution of subject vehicle velocities observed in the naturalistic driving dataset, and runs $N$ separate simulations where the behavior of the target vehicle is sampled based on the driving policy $f(s,a|\Lambda)$. The algorithm outputs estimate of $\Lambda^*$, which is subsequently used to construct the final driving policy to be used in estimation of the rare event probability $p^{IS}_{\epsilon}$ based on Algorithm \ref{algo:is}.
\begin{algorithm}[h]
\SetAlgoLined
\KwResult{$\Lambda^{bid_{max}}_{max}$}
 $[\Lambda_{max}^{B1},..,\Lambda_{max}^{B8}] \leftarrow \text{init\_}\Lambda()$;\\
 \ForEach{$bid \in [B1,..,B8]$}{%
      $p_{max}^{bid} \leftarrow$ simulate\_scene($\Lambda_{max}^{bid},N$) }
 \While{$i < I_{max}$}{
  $bid \leftarrow \text{weighted\_sample(}[p_{max}^{B1},..,p_{max}^{B8}]\text{)}$;\\
  $p_{max} \leftarrow \text{max}([p_{max}^{B1},..,p_{max}^{B8}])$;\\
  \If{$\exp{((p_{max}^{bid} - p_{max})/T_{out})} < $random(0,1)}{
   $j \leftarrow 0$\;
   \While{$j < J_{max}$}{
   $\Lambda \leftarrow \text{sample}(bid)$;\\
   $p\_\epsilon \leftarrow$ simulate\_scene($\Lambda,N$);\\
   \If{$p\_\epsilon > p_{max}^{bid} ${\normalfont or}$  \exp{((p\_\epsilon - p_{max}^{bid})/T_{inn})} < $random(0,1)}{
   $p_{max}^{bid} \leftarrow p\_\epsilon$\\
   $\Lambda_{max}^{bid} \leftarrow \Lambda$\\
   }
   $T_{inn} \leftarrow \text{temperature}(j)$
   }
   }
   $T_{out} \leftarrow \text{temperature}(i)$\\
   $bid_{max} \leftarrow \argmax\limits_{bid}p^{bid}_{max}$\\
 }
 \caption{Simulated Annealing (SA) based optimization procedure for $\Lambda^{*}$}
 \label{algo:br_opt}
\end{algorithm}

\subsection{Situation generation}
\label{sec:scene_gen}

While the bounded rationality model can be used to provide a point estimate of the probability of rare events, the model can also be used to sample new situations of interest to evaluate the performance of the planner under specific circumstances. One simple way to achieve that is by sampling behaviors of other vehicles from the driving policy conditioned on a behavior category. For example, to generate situations of high speed cut-ins at close distances (B1,B2), behaviors can be sampled from the distribution $f(s,a|\Lambda_{\text{B1,B2}})$, where $\Lambda_{B1,B2}$ is the domain of $\Lambda$ after applying the constraints of the respective behavior categories (B1,B2) based on Table \ref{tab:beh_cat}. Although this approach can sample a wide range of behaviors, a more effective technique can use a data-driven strategy consisting of the following steps: (i) acquisition of naturalistic driving data for the scenario under evaluation, (ii) fitting a behavior model based on the data, and (iii) using the behavior model to sample new situations that are not present in the dataset. In this section, we propose an approach to achieve the above objectives.\\
\begin{figure*}[h]
     \begin{subfigure}[b]{.33\textwidth}
         \includegraphics[width=\textwidth]{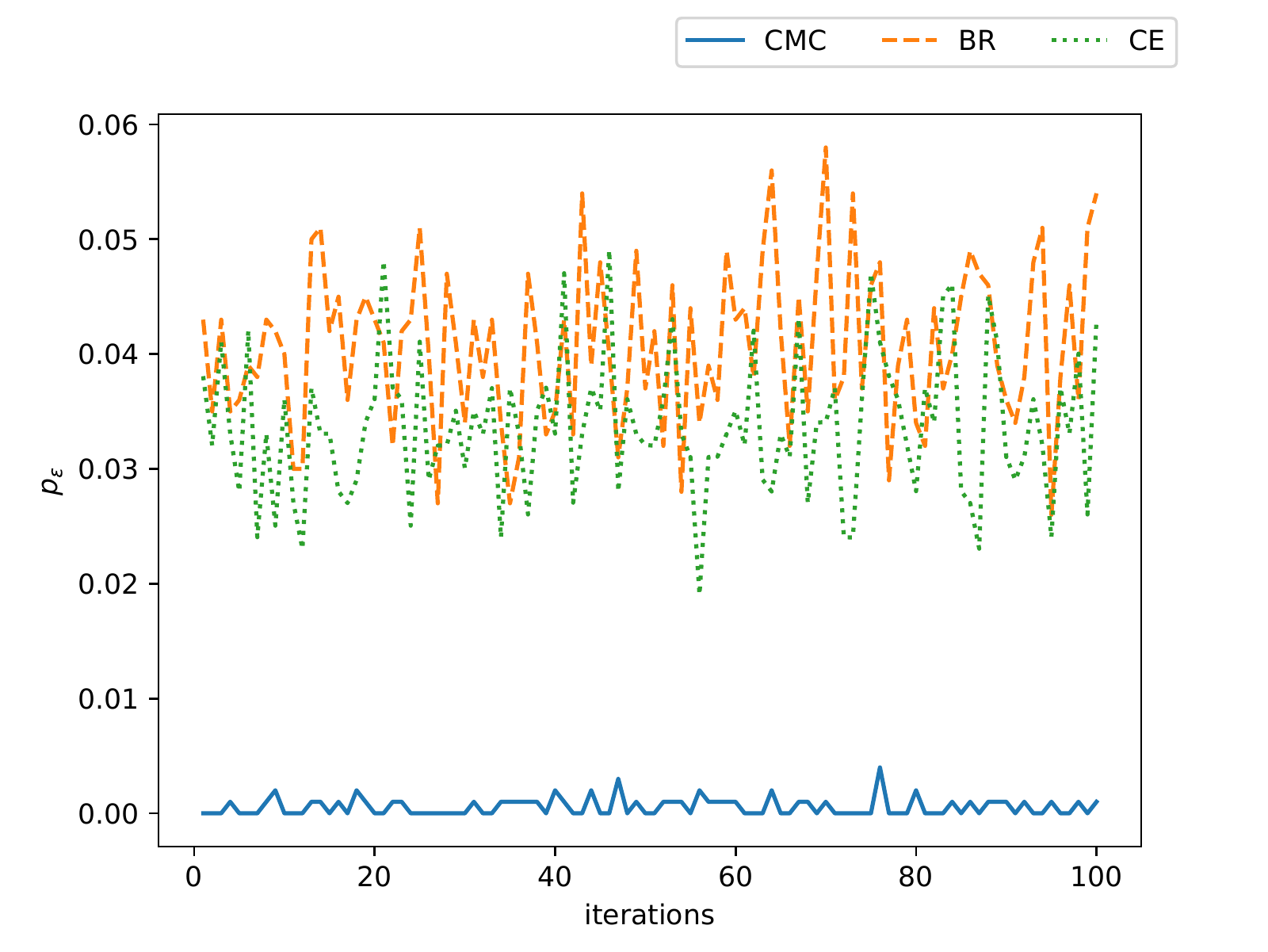}
         \caption{}
         \label{fig:se_iters}
     \end{subfigure}
     \begin{subfigure}[b]{.33\textwidth}
         \includegraphics[width=\textwidth]{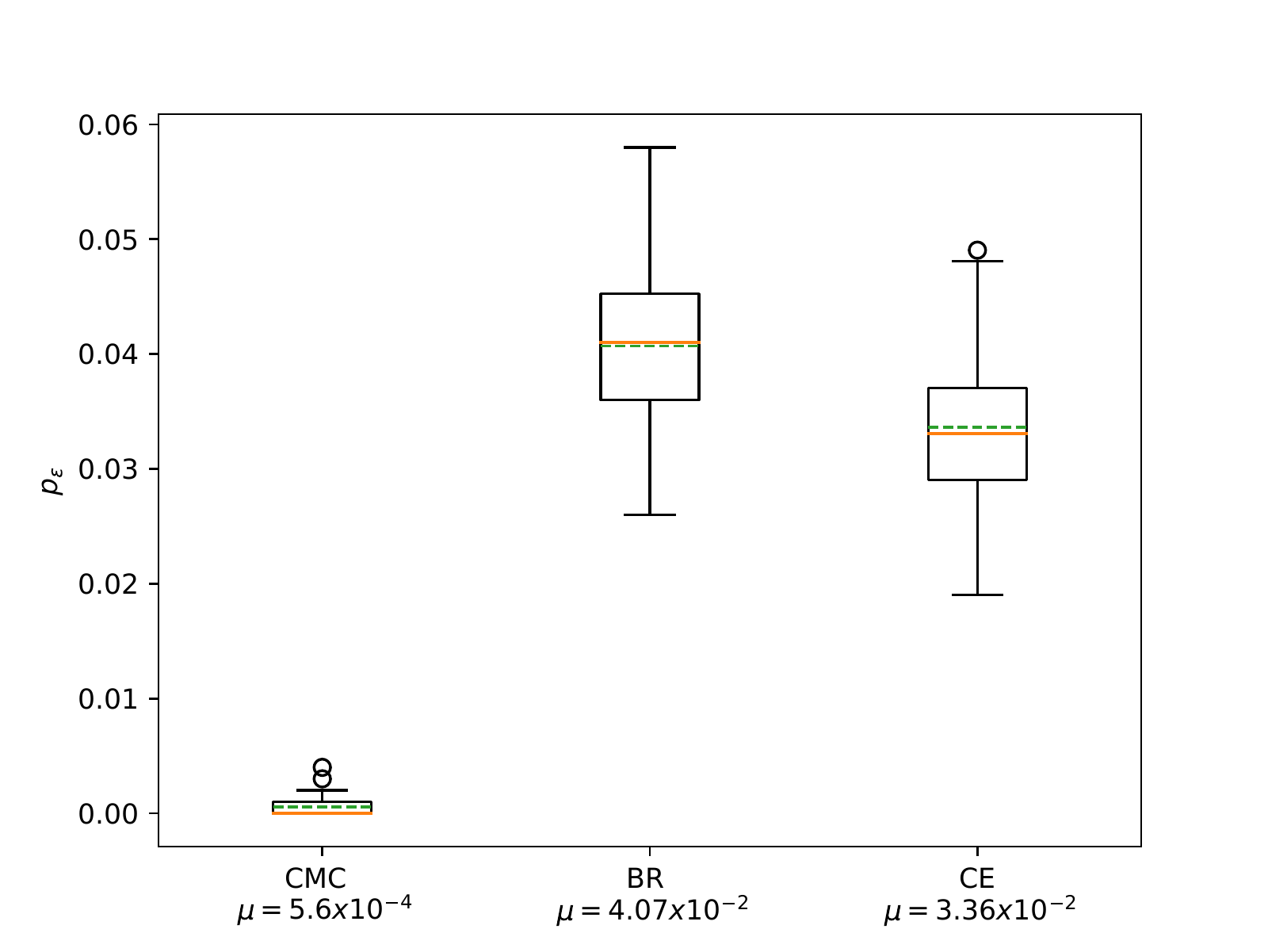}
         \caption{}
         \label{fig:se_var}
     \end{subfigure}
     \begin{subfigure}[b]{.33\textwidth}
        \includegraphics[width=\textwidth]{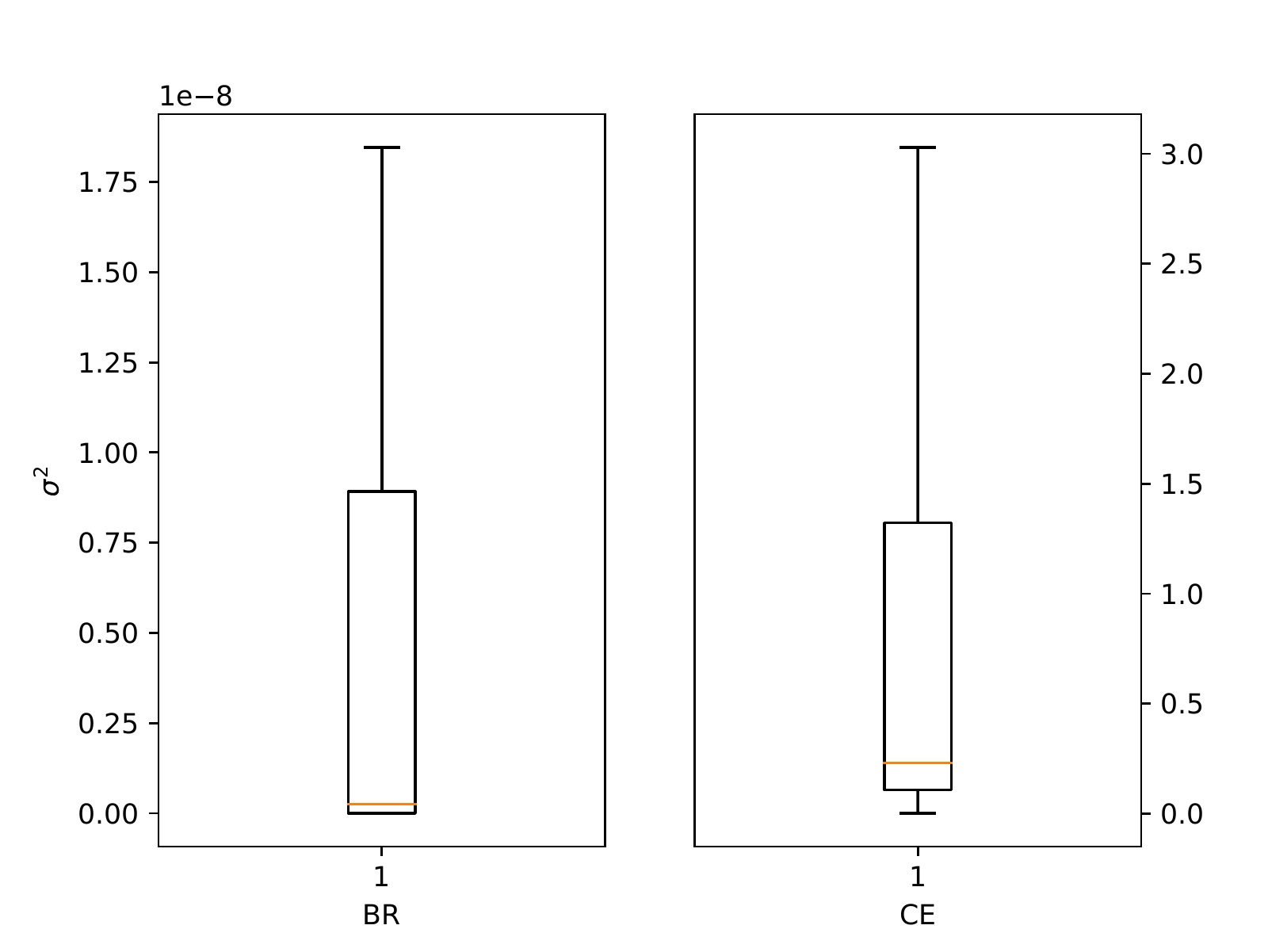}
        \caption{}
        \label{fig:box_plot_w_var}
     \end{subfigure}
     \caption{Comparison based on variance and probability of rare events for the three approaches after 100 simulation runs. (a) probability of rare events ($p_{\epsilon}$ for the three approaches (BR,CE,CMC). x-axis: simulation run, y-axis: probability of rare event corresponding to the run), (b) Box plot of the probability of rare events based on 100 iterations, (c) Box plot of the variance of the likelihood ratio $w$ for BR and CE after 100x1000 runs. Dotted and straight line shows the median and mean values respectively for box plots.}
     \label{fig:se}
\end{figure*}
Compared to the problem of rare event sampling, where we optimize for a single value of $\Lambda$ that maximizes the rare event probability, fitting the model to naturalistic data poses additional challenges. Naturalistic data are often multimodal in nature, i.e, they contain a a mix of different driving behaviors, and thus, a model fitted with a single value of $\Lambda$ cannot capture the variation adequately. In order to resolve this problem, we apply insights from the behavior categories developed earlier, and extend the bounded rationality based model for the more general setting of modeling mixed behaviors. We introduce three mixing parameters $A = \{\alpha_{\delta},\alpha_{\tau},\alpha_{p}\} \in \mathbf{R}_{[0,1]}^{3}$, one for each utility, and correspondingly extend the driving policy formulation of \eqref{eqn:drive_pol_f} to 
\begin{equation}
\begin{split}
    f(s,a|\Lambda,A) = \frac{1}{k}\sum\limits_{i=1}^{k} \alpha_{i}p(s,a|(\lambda_{i}^{+},u_{i})) + \\ (1-\alpha_{i})p(s,a|(\lambda_{i}^{-},u_{i}))
    \label{eqn:f_fit}
\end{split}
\end{equation}
\begin{wrapfigure}{r}{0.2\textwidth}
\setlength{\intextsep}{0pt}%
\setlength{\columnsep}{0pt}%
  \begin{center}
    \includegraphics[width=0.2\textwidth]{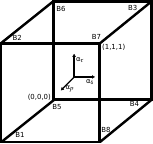}
  \end{center}
  \caption{Visualization of the behavior model that mixes different behaviors categories (B1-B8) based on the mixing parameter $A = {\alpha_{\delta},\alpha_{\tau},\alpha_{p}}$}
  \label{fig:cube_category}
\end{wrapfigure}
where ${\lambda^{+},\lambda^{-}}$ are the positive and negative constraints on the values of the parameter. The distribution in the above model has multiple peaks, and thus allows for mixing multiple behaviors [B1-B8]. A convenient way to visualize this is with a unit hypercube (Figure \ref{fig:cube_category}) where the corners are the behavior categories from Table \ref{tab:beh_cat}, and the parameter $A$ controls the corresponding mix of behaviors. To fit the nine parameters ($\lambda_{\delta}^{+,-},\lambda_{\tau}^{+,-},\lambda_{p}^{+,-},\alpha_{\delta},\alpha_{\tau},\alpha_{p}$) to the observed data, we use least-squares optimization based on Trust Region Reflective algorithm since the parameters are bounded \cite{li1993centering}. Finally, based on the fitted parameters, we use the behavior model of \eqref{eqn:f_fit} to sample new situations that are unseen in the original dataset.

\section{EXPERIMENTS}

In order to evaluate the bounded rationality based behavior model for rare event sampling as well as situation generation, we use the University of Michigan SPMD (Safety Pilot Model Deployment) dataset. SPMD is one of the largest naturalistic driving dataset that was collected over two years, with 2842 equipped vehicles driving a total of 34.9 million miles. Part of the dataset with two months of driving data is publicly available in \cite{spmddataset}, and contains information recorded from the vehicle's data acquisition systems, such as MobilEye camera, CAN bus, and GPS. We follow the approach in \cite{zhao2017trafficnet} to extract 74,449 cut-in events recorded in the dataset, as well as the target and subject vehicle trajectory for 5 seconds immediately following the event. For our experiments, we define rare event as near-crash situations where distance gap between the subject and target vehicle is .01 meter or less and the subject vehicle is not stopped.\\
As a part of the evaluation, we address two specific research questions based on the approaches developed earlier:\footnote{Our code is available at \url{https://bit.ly/2H83i1o}}
\begin{itemize}
    \item RQ1: How does behavior-driven RE sampling compare to crude Monte Carlo and cross-entropy based methods?
    \item RQ2: How well does the bounded-rationality model fit the observed naturalistic driving data?
\end{itemize}

\emph{RQ1:} Based on the theory presented in Section \ref{sec:re_sampling}, we revisit that any system input sampled from the optimal proposal distribution $q^{*} = p(g|\epsilon)$ always leads the system to rare event. A factor that helps judge the goodness of a proposal distribution is the probability of generation of rare events; the general intuition being that a higher probability is an indication of the distribution being closer to $p(g|\epsilon)$. Thus, we first compare the bounded rationality based model developed in the paper (BR) with Cross Entropy (CE) based proposal distribution, as well as baseline crude Monte Carlo sampling (CMC) on the basis of probability of generation of rare events.

As shown in Figure \ref{fig:se_iters}, the proposal distribution based on bounded rationality model outperforms cross-entropy based model consistently across 100 iterations, where each iteration consists of 1000 simulation runs sampled from the respective distributions. The mean ($\mu$) and box plot of the runs is shown in Figure \ref{fig:se_var}. As expected, the both proposal distributions ($\mu = 4.07 \times 10^{-2}$ and $\mu = 3.36 \times 10^{-2}$ for CE and BR respectively) have higher probability of generating rare events compared to crude Monte Carlo sampling ($\mu = 5.6 \times 10^{-4}$). \\
Along with the probability of occurrence of rare events, another important metric to evaluate the quality of the proposal distribution is the variance of the likelihood ratio $w$ (line 5 in Algorithm. \ref{algo:is}). A variance closer to 0 is an indication of the proposal distribution being closer to the optimal distribution ($p(g|\epsilon)$), which is a zero variance distribution. A lower variance also reduces the width of the confidence intervals of the $p_{\epsilon}^{IS}$ estimates, thus reducing the relative error between the estimate and the true probability $p_\epsilon$. Figure \ref{fig:box_plot_w_var} compares the variance of the weights that were sampled from the respective proposal distribution for BR and CE. Variance was calculated for the 1000 runs in each iteration, and the box plot shows their dispersion over 100 iterations. As seen from the figure, the variance is significantly lower for BR compared to CE, indicating that BR provides a more accurate estimate of $p_{\epsilon}$ compared to CE. Thus, the results show that categorizing based on the behaviors, and  optimizing over them to find a crash-prone driving policy can act as an effective rare event sampling strategy.\\
\begin{figure*}[h]
     \centering
     \begin{subfigure}[b]{0.35\textwidth}
         \includegraphics[width=\textwidth]{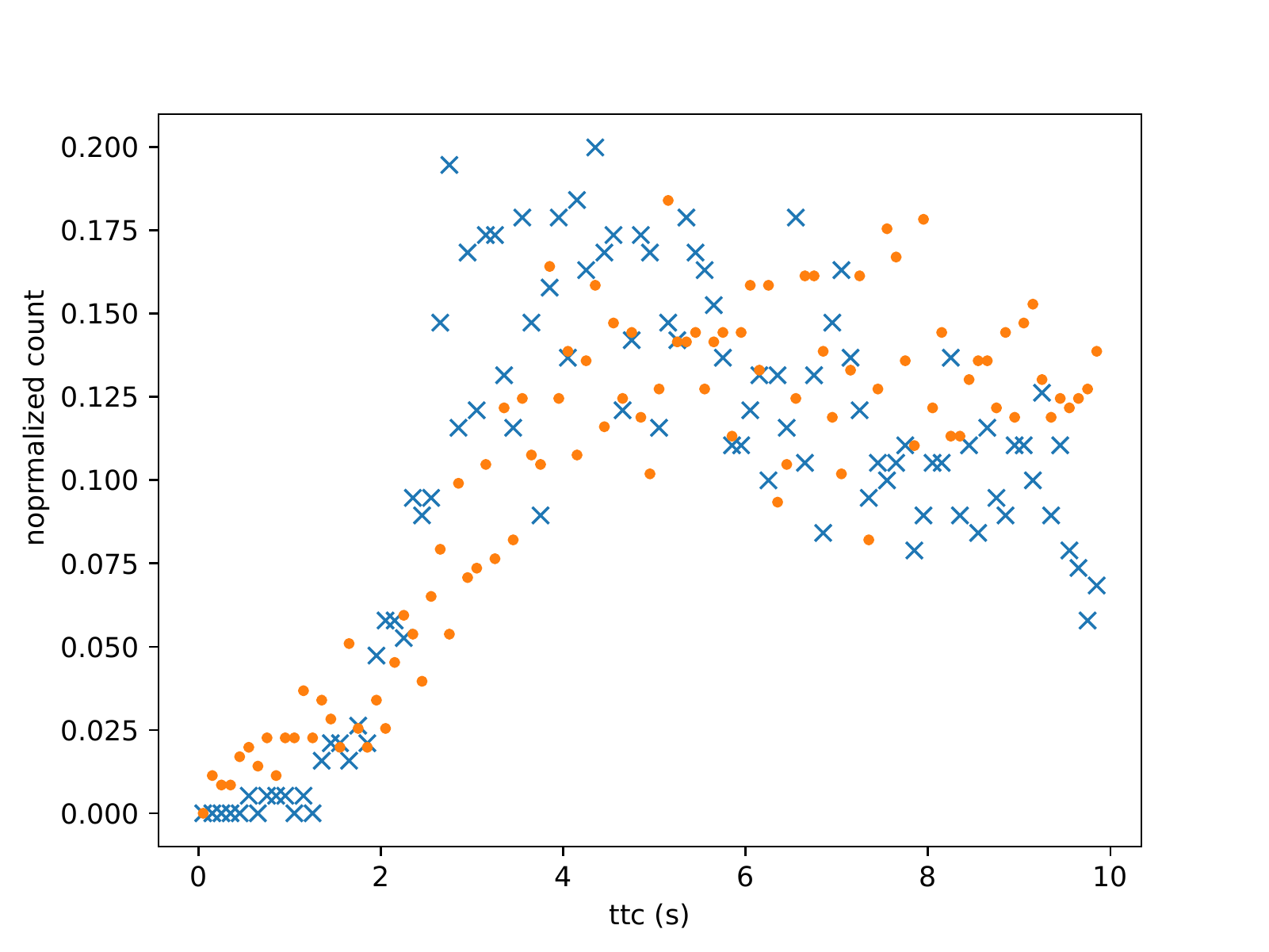}
         \caption{}
        \label{fig:ttc_gen_comp}
     \end{subfigure}
     \begin{subfigure}[b]{0.35\textwidth}
         \includegraphics[width=\textwidth]{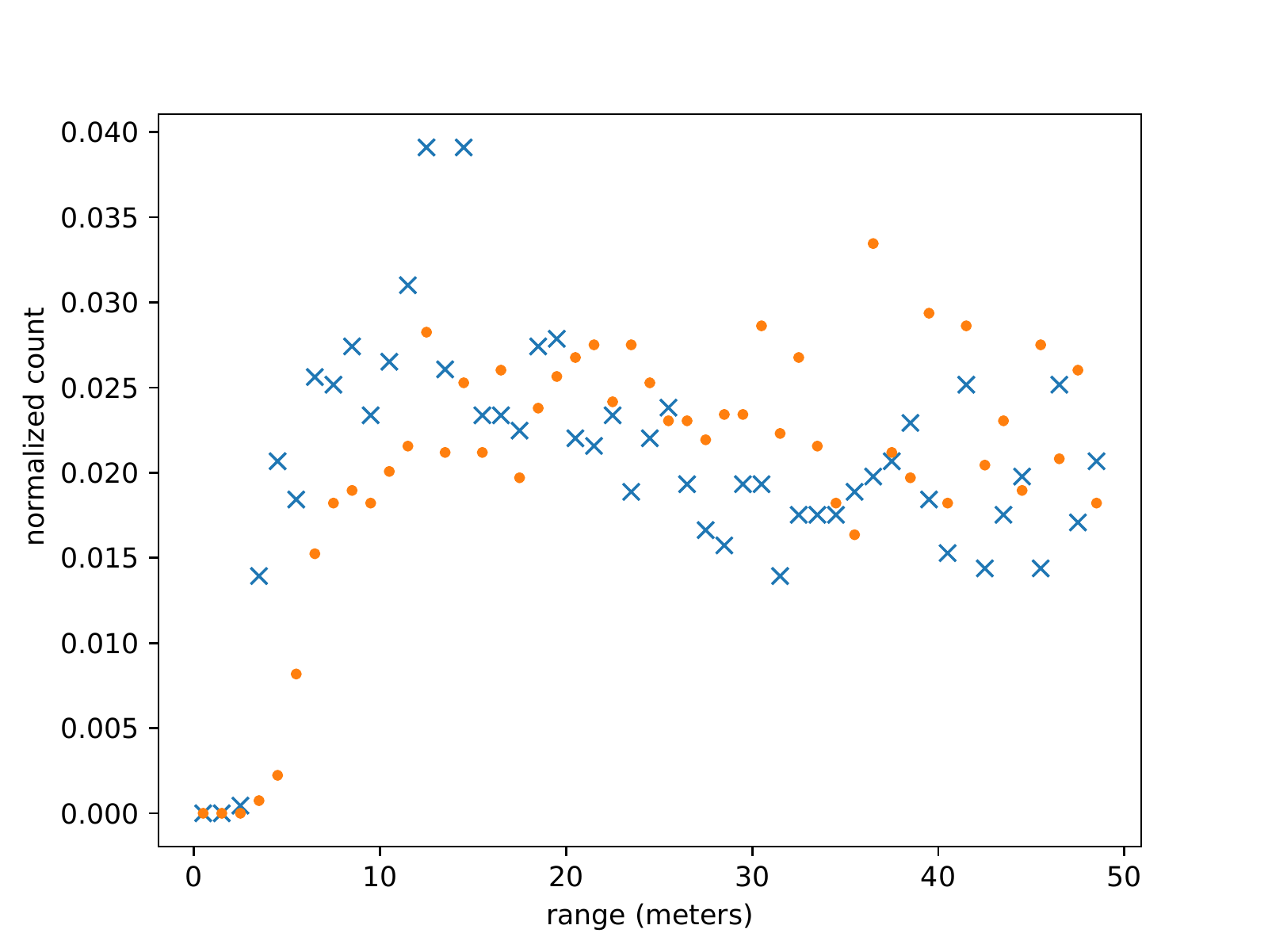}
         \caption{}
         \label{fig:range_gen_comp}
     \end{subfigure}
     \begin{subfigure}[b]{0.25\textwidth}
         \includegraphics[width=\textwidth]{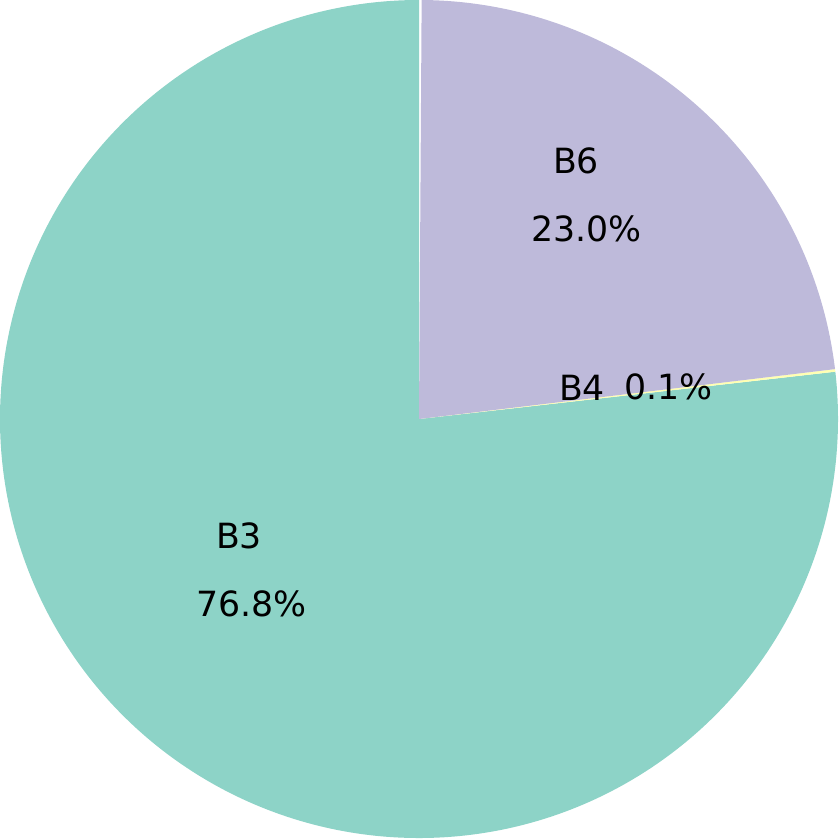}
         \caption{}
         \label{fig:b_pie}
     \end{subfigure}
     \caption{(a,b): Comparison of generated (blue cross) and naturalistic data (orange dots) for low speed cut-in situations (subject vehicle speed is less than 15 meters per second). x-axis: metric values (a: ttc (secs), b: range (meters)), y-axis: probability. (c) Distribution of behaviors for all situations in the dataset.}
     \label{fig:data_gen}
\end{figure*}
\begin{figure}[h]
     \centering
     \begin{subfigure}[b]{0.35\textwidth}
         \centering
         \includegraphics[width=\textwidth]{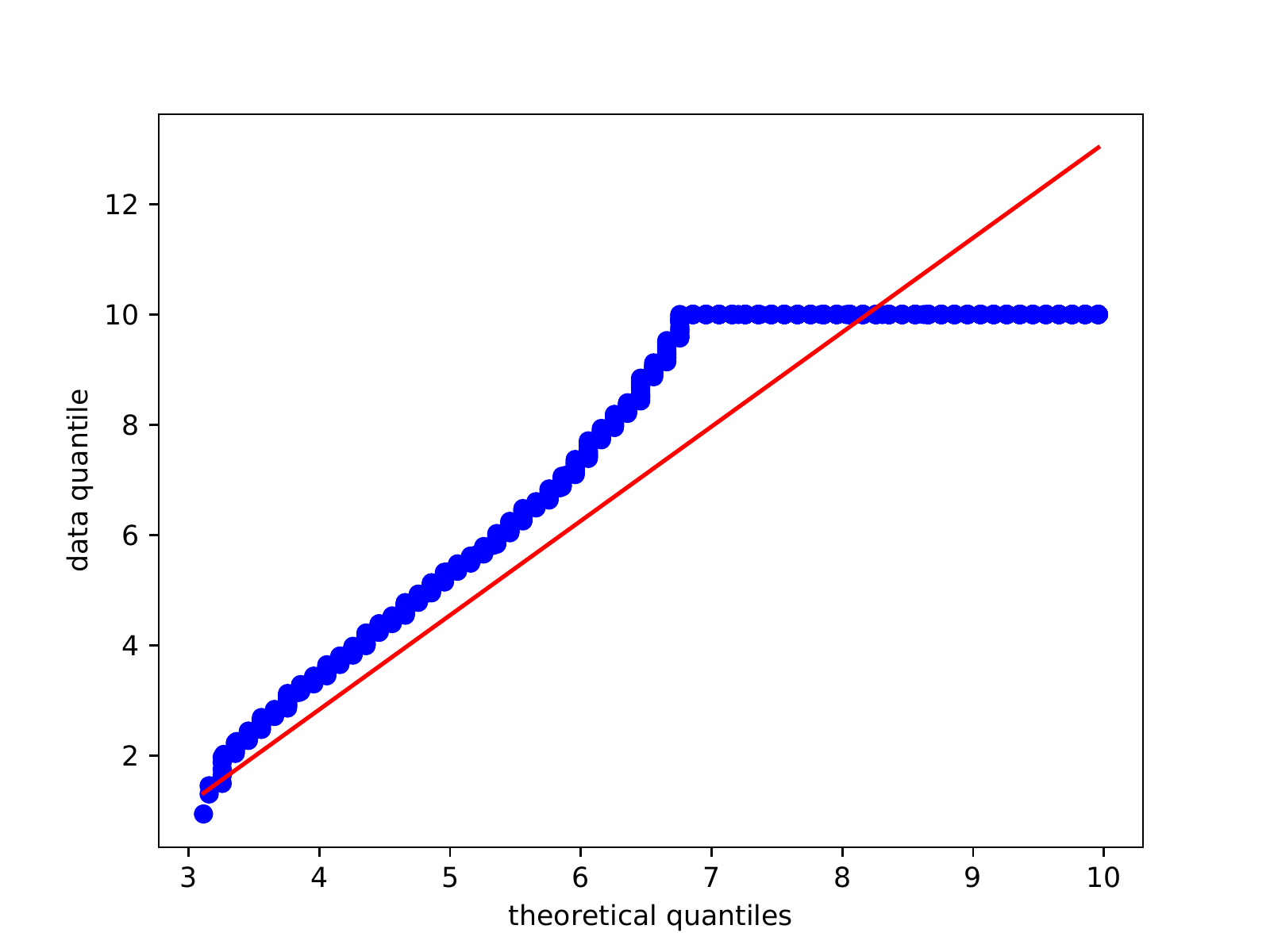}
         \caption{}
         \label{fig:ttc_qq}
     \end{subfigure}\\
     \begin{subfigure}[b]{0.35\textwidth}
         \centering
         \includegraphics[width=\textwidth]{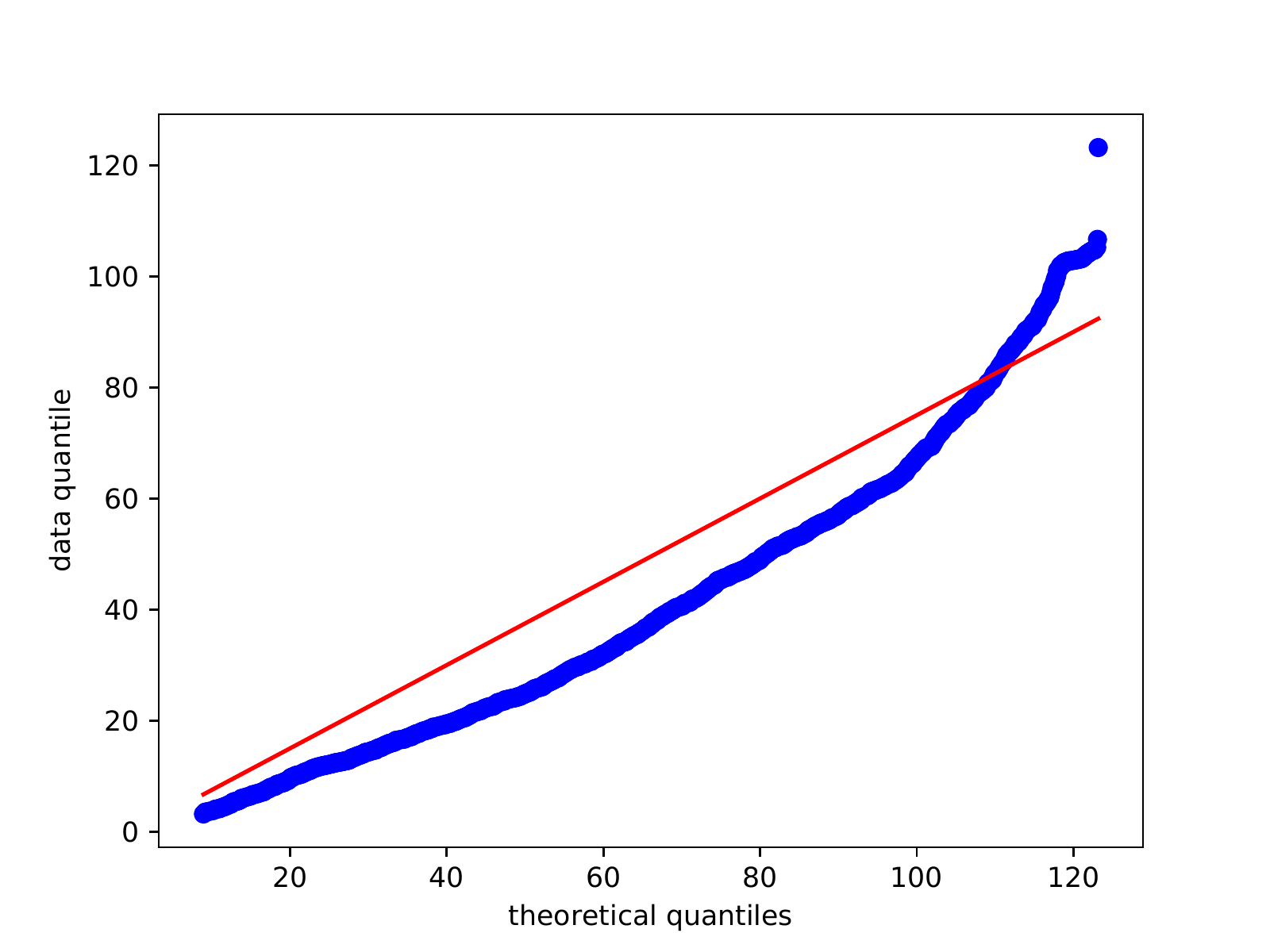}
         \caption{}
         \label{fig:range_qq}
     \end{subfigure}
     \caption{QQ plot for naturalistic data with respect to bounded rationality based behavior model (a) ttc (b) distance-gap. x-axis: theoretical quantiles of the distribution for the behavior model, y-axis: empirical quantiles from naturalistic data.}
     \label{fig:qq}
\end{figure}
\emph{RQ2:} As a part of the second research question, we evaluate how well bounded rationality based behavioral model can model naturalistic traffic data. To that end, we perform a random (80-20) split of the SPMD lane change dataset. We use 80$\%$ of the data to fit the parameters based on the approach discussed in Section \ref{sec:scene_gen}, and use the remaining for evaluation. Since the distributions are dependent on the speed, we repeat our analysis for low ($\leq 15\ \text{m/s}$), medium ($15 - 25\ \text{m/s}$), and high speed ($>25\ \text{m/s}$) situations. Figure \ref{fig:data_gen} shows a visual comparison of the range and time-to-collision distribution of the generated and observed data for low speed situations, shown as blue crosses and orange dots, respectively. To evaluate the fit analytically, the QQ plots are shown in Figures \ref{fig:ttc_qq},\ref{fig:range_qq}. QQ plots are an effective tool to measure the fit of observed data to a theoretical distribution. The x-axis represents the quantiles of the distribution, the y-axis represents the observed data quantiles, and the blue dots maps a quantile from the data to the distribution. As seen in the figure, the strong linear relation (Pearson correlation coefficient $\rho=.919,.98$ resp. for ttc and distance gap) between the two indicate that the observed data from the dataset is distributed according to the fitted bounded rationality based model.

\section{RELATED WORK}

\emph{rare event sampling:} There is significant variation in the approaches used for rare event sampling, including importance sampling, subset simulation, and splitting \cite{beck2017rare}. Subset simulation has been used in domains where the rare event probability $p_{\epsilon}$ can be expressed as a product of factors of higher probability, and the approach estimates the factor probabilities separately. Splitting is a related technique where the simulation makes iterative copies of the system state that leads to a state close to a rare event, and subsequently running simulations from that state. rare event sampling also has a rich history of application in various domains such as aerospace \cite{blom2009rare}, systems biology \cite{sandmann2009rare}, telecommunication \cite{blanchet2007rare}. Blom et al. \cite{blom2009rare} apply splitting technique to the problem of safety verification for air traffic control in order to avoid rare events such as aircraft collision. Blanchet and Mandjes \cite{blanchet2007rare} apply an importance sampling based technique for queuing systems, and highlight the relevance of standard deviation of the estimate in a good IS proposal distribution. In the domain of autonomous vehicles, Zhao et al. use rare event sampling for accelerated evaluation of AV for lane change scenarios \cite{zhao2017accelerated}. Kelly et al. use rare event sampling for testing driving policies that are based on end-to-end learning \cite{o2018scalable}. Both approaches use cross-entropy based importance sampling as the simulation technique. To the best of our knowledge, we present the first approach that highlights the importance of different driving behaviors for rare event sampling in the context of autonomous vehicles.\\
\emph{Behavior modeling:} Most previous approaches to modelling traffic behavior are limited to the deterministic case, where the behavior of vehicles were modelled as differential equations. Examples of such models include the Intelligent Driver Model \cite{treiber2000congested}, along with its extensions such as the Newell car-following model \cite{newell2002simplified}. When applied to the problem of ADS simulation, these approaches are limited in their ability to model the variation of human traffic behavior, including positive, negative, and edge case behavior. In the broader field of behavior modeling, there is extensive body of literature on modeling and simulation of pedestrian behavior under varied situations. Popular approaches use variations of the Social Forces Model (SFM) \cite{helbing2000simulating}, where the behavior of agents is modeled as a dynamical system containing attractive and repulsive forces, and the final behavior is the result of all such forces acting on the agent. Although SFM provides an intuitive modeling paradigm to model agent movement, it has been shown to be difficult to calibrate the models to real empirical data due to the forces not being linearly additive in nature. To address the shortcomings of the social forces model, potential-based methods follow an agent-free model, where the behavior is not modeled individually for every agent like in SFM. Instead, potential based methods treats goals and obstructions as a continuous potential field, and the resulting behavior is the solution to energy minimization problem in the field. Potential field based methods can be considered to be a special case of utility-based methods. However, like most utility based methods, potential field models work under the assumption that the behavior always follow the optimal path. We consider this assumption restrictive, and address this by using bounded rationality in our approach. Yang and Peng \cite{yang2010development} develop an errable driver model to model sub-optimal driving behaviors, including distraction and perceptual errors. The model measures the stochastic error in driving decisions based on the specified error factors. Compared to the errable model, our approach is based on a more general utility-driven framework, and thus can be applied to a wider variety of driving situations. 

\section{CONCLUSION}
In this paper, we develop a novel driving behavior model based on the theory of bounded rationality to model traffic behavior for evaluation of autonomous vehicles. We apply the behavior model to two cases: (i) generation of rare event situations and estimating the probability of rare events, and (ii) applying the behavior model to generate new synthetic data for testing behavior planners. We evaluate our proposed model based on a large naturalistic dataset and show that bounded rationality based behavior model can improve on crude Monte Carlo sampling by an order of $10^4$, and compared to cross-entropy sampling, it provides $33\%$ speedup and $99\%$ reduction in variance. We also show that synthetic data sampled from the developed behavior model has strong correlation to naturalistic driving data.

\addtolength{\textheight}{-12cm}   





\bibliographystyle{IEEEtran}
\bibliography{main}

\end{document}